\documentclass[english,nofootinbib,twocolumn,rmp,aps]{revtex4-1}
\usepackage{graphicx}
\usepackage{dcolumn}
\usepackage{bm}
\usepackage{hyperref}

\renewcommand{\vec}[1]{{\mathbf{#1}}}

\newcommand{\beq}{\begin{eqnarray}}
\newcommand{\eeq}{\end{eqnarray}}

\renewcommand{\vec}[1]{{\mathbf{#1}}}

\usepackage{amsmath,amssymb,amsfonts}
\usepackage{braket}
\usepackage{upgreek}

\newcommand{\pa}{\left(\frac{\partial}{\partial z}\right)^a}

\newcommand{\lap} {\triangle}

\newcommand{\CC}{{\mathbb{C}}}
\newcommand{\Ly}[1]{\lap_x #1 + \frac{a}{y} #1_y + #1_{yy}}

\newcommand{\R}{\mathbb R}

\renewcommand{\vec}[1]{\boldsymbol{#1}}

\def\a{\alpha}

\newcommand{\pab}{\left(\frac{\partial}{\partial \bar z}\right)^a}

\begin{document}
\title{ Fractional Electromagnetism in Quantum Matter and High-Energy Physics}
\author{Gabriele La Nave}\affiliation{Department of Mathematics,
University of Illinois, Urbana, IL 61801, U.S.A.}
\author{Kridsanaphong Limtragool}
\affiliation{Department of Physics, Faculty of Science, Mahasarakham University, Khamriang Sub-District, Kantharawichai District, Maha-Sarakham 44150, Thailand}
\author{Philip W. Phillips}
\affiliation{Department of Physics,
University of Illinois
1110 W. Green Street, Urbana, IL 61801, U.S.A.}
\date{\today}

\begin{abstract}

We present here a theory of fractional electricity and magnetism which is capable of describing phenomenon as disparate as the non-locality of the Pippard kernel in superconductivity and anomalous dimensions for conserved currents  in holographic dilatonic models.  While it is a standard result in field theory that the scaling dimension of conserved currents and their associated gauge fields are determined strictly by dimensional analysis and hence cannot change under any amount of renormalization, it is also the case that the standard conservation laws for currents, $dJ=0$, remain unchanged in form  if any differential operator that commutes with the total exterior derivative, $[d,\hat Y]=0$, multiplies the current.   Such an operator, effectively changing the dimension of the current, increases the  allowable gauge transformations in electromagnetism and is at the heart of N\"other's  second theorem.  However, this observation has not been exploited to generate new electromagnetisms.  Here we develop a consistent theory of electromagnetism that exploits this hidden redundancy in which the standard gauge symmetry in electromagnetism is modified by the rotationally invariant operator, the fractional Laplacian. We show that the resultant theories all allow for anomalous (non-traditional) scaling dimensions of the gauge field and the associated current.  Using the Caffarelli/Silvestre\cite{CS2007} theorem, its extension\cite{csforms} to p-forms and the membrane paradigm, we show that either the boundary (UV) or horizon (IR) theory of holographic dilatonic models are both described by such fractional electromagnetic theories.   We also show that the non-local Pippard kernel introduced to solve the problem of the Meissner effect in elemental superconductors can also be formulated as a special case of fractional electromagnetism.  Because the holographic dilatonic models produce boundary theories that are equivalent to those arising from a bulk theory with a massive gauge field along the radial direction, the common thread linking both of these problems is the breaking of $U(1)$ symmetry down to ${\mathbb Z}_2$.   We show that the standard charge quantization rules fail when the gauge field acquires an anomalous dimension.  The breakdown of charge quantization is discussed extensively in terms of the experimentally measurable modified Aharonov-Bohm effect in the strange metal phase of the cuprate superconductors.

\end{abstract}

\maketitle
\tableofcontents

\section{Introduction}

An unexpected theoretical consequence of the Faraday induction experiment is that a theory of electricity and magnetism in terms of independent electric and magnetic fields is redundant.  This redundancy is alleviated by  formulating electromagnetism as a gauge theory in which the basic building block is the gauge field $A$ and the field strength $F=dA$, where $d$ is the total exterior derivative.  The individual magnetic and electric fields are determined by various components of the field strength, $F_{0i}=-E_i$ and 
$F_{\mu\nu}=\epsilon^{\mu\nu\kappa}B_{\kappa}$.  The redundancy between $E$ and $B$  is now expressed as the gauge invariant condition 
\beq\label{eq:U1transformation}
A_\mu\rightarrow A_\mu+\partial_\mu\Lambda(x),
\eeq
where $\Lambda(x)$ is a phase angle.  Since $\Lambda$ is dimensionless, the engineering dimension of $A$ is unity.  This has a fundamental consequence in field theory.  As it is the gauge field that serves as the source for the conserved 4-current, $J_\mu$, thereby entering the action in the form $J_\mu A^\mu$, the dimension of $J_\mu$ is determined solely by the volume factor in the action.   Indeed,  a standard homework problem in field theory is to prove that conserved currents cannot acquire anomalous dimensions under renormalization as long as gauge invariance remains intact\cite{gross,peskin}.  The argument provided by Gross\cite{gross} gets at the main point.  It is the commutator of the charge density with any $U(1)$ field, $\phi(x)$,
\beq
\delta(x_0-y_0)[J_0(x),\phi(y)]=\delta \phi(y)\delta^d(x-y),
\eeq 
that fixes the scaling dimension of the conserved current.  Here $\delta \phi(y)$ is the change in the field $\phi$ to linear order upon acting with the  $U(1)$ transformation and $J_0$ is the charge density.   Consequently, that $[J_\mu]=d-1$ is sacrosanct.  Note the covariant derivative,
heuristically written as $D-iqA$, only fixes the dimension of the product
$[qA]=1$.  Hence, it is entirely possible to construct theories\cite{wise}
in which $q$ and $A$ have arbitrary dimensions still within the confines of
the standard framework.  

\begin{figure}
	\includegraphics[scale=0.5]{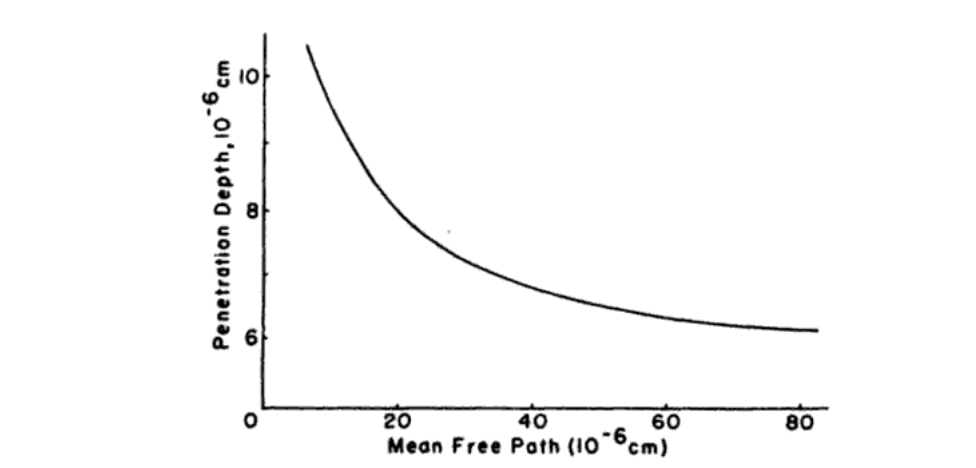} 
    \caption{Variation of the penetration depth of a superconductor with the mean-free path as determined by the degree of alloying Sn with Indium impurities.  (Reprinted from B. Pippard, Proc. Roy. Soc. (London) A216, 547 (1953)). } \label{pippard}
\end{figure}

However, as the work of Brian Pippard\cite{pippardref} underscores, the rules governing the interaction of light and matter are not set in stone but rather are emergent.  Consequently, the physics might dictate a deviation from the scaling of the current derived in the opening paragraph.   To illustrate how this plays out, consider the London\cite{London} relationship 
\beq\label{london}
J_s=-\frac{4\pi n_s e^2}{m c^2}A=-A/\lambda^2
\eeq
which was instrumental in solving the Meissner problem in a superconductor.  Here $J_s$ is the current inside a superconductor, $A$ the gauge field,  $n_s$ the superfluid density, $e$ the electric charge and $m$ the bare electron mass.  While this equation is simple and follows from basic quantum mechanics, it is its simplicity that admits an immediate problem.  Namely, the penetration depth, $\lambda$ depends only on fundamental constants and intrinsic properties of the condensed state. Further, all that is required to specify the current at a single point is the value of the gauge field at that point.  That is, the current and the gauge field  are related in a point-like local fashion.  It is precisely this idea of locality that Pippard was interested in testing.   He\cite{pippardref} found experimentally (see Fig. (\ref{pippard})) in samples of Sn that increasing the Indium content by as little as 3$\%$ led to a drastic increase in the $T=0$ penetration depth with no change in the superconducting transition temperature $T_c$ or any other thermodynamic property including  $n_s$ and $m$ which appear explicitly in London's theory.  Pippard reasoned that since the mean-free path was affected by the impurities, the current must be related to the gauge field on distances at least as large as the mean-free path.  Consequently, the London equation must be invalid and he proposed instead,  in collaboration with R. G. Chambers, that the current and the gauge field are {\it in fact} conjoined in an explicitly non-local,
\beq
J_s(\vec r)=-\frac{3}{4\pi c\xi_0\lambda}\int\frac{(\vec R(\vec R\cdot\vec A (\vec r')) e^{-( R/\xi(\ell))}}{ R^4} d^3\vec r',
\eeq
 relationship with $\vec R=\vec r-\vec r'$.  Here $\xi(\ell)$ is a length determined by the mean-free path, $\xi_0$ is a constant having dimensions of length, and the integral is over the entire volume of the metal. It is from the explicit integration over all space that the non-locality is manifest.  What Pippard\cite{pippardref} found is that this relationship better described his data than does Eq. (\ref{london}) and hence advocated that currents in superconductors are inherently non-local in terms of their dependence on the applied field.  This appears to be the first time such an idea of non-local currents was advocated, a theme which will help demystify our analysis of the strange metal.

To put the Pippard kernel in a broader context, we appeal to the argument by Weinberg\cite{weinberg,wittensup}.  In a superconductor, the local $U(1)$ symmetry is broken to $\mathbb Z_2$.  Consequently, it is the coset group $U(1)/{\mathbb Z}_2$ that parameterizes the phase such that $\phi$ and $\phi +\pi\hbar c/e$ are equivalent. Because of this spontaneous symmetry breaking, the superconducting phase $\phi$ becomes rigid, i.e, the phase stiffness or the superfluid density is nonzero. This results in a superconducting state in which the matter Lagrangian can be  expanded around the minimum at which $\nabla\phi - A=0$.   Retaining only the terms to second order leads to a quadratic action,
\beq\label{matterLag}
L_m &=& L_{m0} - \frac{1}{2}\int d\vec xd\vec x' C^{\mu \nu}(\vec x,\vec x')\, \left( \vec A_\mu (\vec x)-\partial _\mu \phi(\vec x) \right)\nonumber\\&\times&\left( \vec A_\nu (x')-\partial _\nu \phi(\vec x')\right),
\eeq
around this stable minimum.  Here $ L_{m0}$ is independent of $A$ and $\phi$, and the kernel satisfies the symmetry $\ C^{\mu \nu}(\vec x,\vec x')= C^{\nu \mu}(\vec x,\vec x')$.  Quite generally then, we find that the current in the superconducting state
\beq\label{pipJ}
J_i=-\int \; C_{ij} (\vec x,\vec x')   \left( \vec A_j(\vec x')-\nabla _j'\phi(\vec x')\right) d^3x'
\eeq
is a non-local function of the gauge field, $A(\vec x')$.  The Pippard kernel just amounts to a specific choice for $C_{ij}(\vec x,\vec x')$.  In general, the range or strength of the underlying interactions determines the degree to which the kernel $C_{ij}$ is non-local.  As the physics underlying Eq. (\ref{pipJ}) is beyond that entailed by the Maxwell equation $\nabla\times\nabla\times \vec A=4\pi\vec J/c$, also true of the London equation since both require a mass for the gauge field, a natural question arises:  Is there a general formulation of electromagnetism from which such constitutive relationships arise? 

Possible insight to the answer lies in the dimension of the  current in Eq. (\ref{pipJ}).   Specifically,  from Eq. (\ref{pipJ}) the units of the current are $d-d_C-1$, where $d_C$ is the engineering dimension of the kernel $C_{ij}$ which only reduces to the standard result of $d-1$ if $d_C=0$.   That the current acquires an anomalous dimension as a result of breaking $U(1)$ to ${\mathbb Z}_2$ does not seem to have been pointed out previously. Because the current enters the action in the combination $J_iA^i$, the non-traditional scaling of the current in the case of Eq. (\ref{pipJ}) requires either the presence of  a running coupling constant or a non-traditional scaling of the vector potential.   Since the former would lead to an ill-defined charge in the superconducting state, whereas the charge quantization experimentally is $2e$, the only option is that the gauge field must also have an anomalous dimension.  The natural question that arises is:  is symmetry breaking necessary for non-traditional scaling of the vector potential and the current to obtain?  It is this basic question we investigate here in this Colloquium.  What we do here is show that such non-traditional scaling  necessitates a new form of electromagnetism which entails non-local relationships between currents and gauge fields.  By studying dilatonic holographic models, we will be able to delineate the general mechanism for anomalous dimensions for currents and gauge fields.  In such models, a mass for the gauge field in the IR couples to the current in the UV.  However, the IR lives in the bulk while the UV degrees of freedom are confined to the boundary.  A fractional gauge theory at the boundary captures this dichotomy.  The fractional theory at the boundary contains the fractional Laplacian in which the mass  in the IR (or equivalently the explicit dilaton coupling) determines the power of the fractional Laplacian.  The experimental consequences in terms of observables such as the Aharonov-Bohm phase are discussed in the context of the strange metal in the cuprates.  The lack of charge quantization in fractional theories is also highlighted because one of the standard and crucial facts in the quantization of the EM field is incarnated in the form the action,
\beq
S(\lambda) =\int \; dt\; \frac{1}{2} m \dot{\lambda }^2+ e\int_\lambda A ,
\eeq
 a classical charged particle obeys while moving  along a path $\lambda$ with a vector potential $A$.
The path integral,
\beq
Z=\int \;  \mathcal D \lambda {\rm exp} \left( \frac{ i}{\hslash} S(\lambda)\right),
\eeq
is well defined precisely when the flux $e \,\int _\ell A$ for closed loops $\ell$ is an integral multiple of $h=2\pi \, \hslash$. In other words, one needs the vector potential $A$ (thought of as a differential form) to satisfy the condition that
\beq
\int _\ell e A\in h \mathbb Z
\eeq
for every {\it closed} loop $\lambda$. This condition is what is usually aptly named the integrality condition for the cohomology class of $e A$ to be an i{\it ntegral class}. This is basically {\it quantization of charge} and it turns out to be equivalent to the geometric requirement that the form $F_A=dA$ be indeed the curvature of a connection $D=d-eA$ on a $U(1)$ principal bundle $\mathcal P$ \begin{footnote}{This is of course more significant in Dirac's work where $dA$ is not an exact form, that is when $A$ is singular along a ~2-dimensional submanifold of space-time $M$.}\end{footnote}.  As we will see, when the dimension of $A$ is no longer unity, the issue of quantization of charge becomes subtle.

\section{Preliminaries}

To understand how anomalous dimensions can arise in general, we appeal to a simple argument at the heart of gauge transformations in standard electromagnetism.  We start by writing
the action for the energy density in electricity and magnetism
\beq\label{FEN}
S&=&-\frac{1}{4}\int d^dx F^2\nonumber\\
&=&\frac12\int d^dx A_\mu(x)(\eta^{\mu\nu}\partial_\mu^2-\partial^\mu\partial^\nu)A_\nu(x)
\eeq
in terms of its Fourier components,
\beq\label{Maction}
S&=&\frac12\int \frac{d^dk}{2\pi^d} A_\mu(k)[k^2 \eta^{\mu\nu}-k^\mu k^\nu]A_\nu(k)\nonumber\\
&=&\frac12\int d^dk A_\mu(k)M^{\mu\nu}A_\nu(k).
\eeq
The crucial observation is that the matrix $M^{\mu\nu}$ vanishes whenever $A_\nu(k)=k_\nu\Lambda(k)$ or equivalently, the matrix $M$,
\beq
M^{\mu\nu}k_\nu=0,
\eeq
has a zero eigenvalue.  Consequently, inverting $M$ to obtain the photon propagator is problematic and the origin of the well known Fadeev-Popov\cite{fadeevpopov} gauge fixing trick. Note that  $ik_\nu$ is just the Fourier transform of the local gauge transformation, $\partial_\mu \Lambda$, and as a consequence, the generator of the gauge symmetry determines the form of the  zero eigenvector.  

However, if $k_\nu$ is an eigenvector, then so is $f k_\nu$, where $f$ is a scalar. Whence, there are a whole family of eigenvectors,
\beq\label{mgen}
M_{\mu\nu}f k^\nu=0,
\eeq
that satisfy the zero eigenvalue condition.  Since $f k_\nu$ is the generator of the gauge symmetry, there are some constraints on $f$.  First, $f$ must be rotationally invariant.  Second, it cannot change the fact that $\Lambda$ is dimensionless; equivalently it cannot change the fact that $A$ is a 1-form.   As a result, $f$ cannot be a dimensionful constant such as the density\footnote{Implicit in the Karch and Hartnoll work\cite{hk} is a transformation in which $\Lambda$ acquires dimensions.  Such a transformation is not compatible with a $U(1)$ gauge theory. }.  Such a theory would be inherently coordinate dependent.  Third, $f$ must commute with the total exterior derivative; that is, $[f,k_\mu]=0$.  A form of $f$ that satisfies all of these constraints is $f\equiv f(k^2)$.   In momentum space, $k^2$ is simply the Fourier transform of the Laplacian, $-\Delta$.  As a result, the general form for $f(k^2)$ in real space is just the Laplacian
raised to an arbitrary power and the generalization in Eq. (\ref{mgen}) implies that there are a multitude of possible electromagnetisms that are
 invariant under the transformation,
 \beq
 A_\mu\rightarrow A_\mu + f(k^2)ik_\mu\Lambda,
 \eeq
 or in real space,
  \beq
  A_\mu\rightarrow A_\mu + f(-\Delta) \partial_\mu\Lambda,
  \eeq
  resulting in $[A_\mu]=1+2[f]=\gamma$.  
The standard Maxwell theory is just a special case in which $\gamma=1$.  In general, the theories that result for $\gamma\ne 1$ allow for the current to have an arbitrary dimension not necessarily $d-1$. Consistent with the zero-eigenvalue of $M$, such theories all involve the fractional Laplacian raised to a power and hence transform as 
\beq\label{correcteq}
A_\mu\rightarrow A_\mu+(-\Delta)^{(\gamma-1)/2}\partial_\mu\Lambda,\quad [A_\mu]=\gamma.
\eeq
The definition of the fractional Laplacian we adopt here is due to Reisz:
\beq\label{reisz}
(-\Delta_x)^\gamma f(x)=C_{n,\gamma}\int_{\R^n}\frac{f(x)-f(\xi)}{\mid {x-\xi}\mid^{n+2\gamma}}\;d\xi
\eeq
for some constant $C_{n, \gamma}$.  Note rather just depending on the information of $f(x)$ at a point, the fractional Laplacian requires information everywhere in ${\mathbb R}^n$. 
We will refer to all such theories as having anomalous dimensions.  Because the fractional Laplacian is a non-local operator, the corresponding gauge theories are all non-local and offer a much broader formulation of electricity magnetism than previously thought possible.  All such anomalies can be understood as particular instances of N\"other's Second Theorem and arise naturally from holographic theories with bulk dilaton couplings.   As we will see, such theories have far-reaching experimental consequences and might capture what is strange about the strange metal in the normal state of the cuprates and other non-Fermi liquids in strongly correlated quantum matter.

\subsection{Prior Fractional Electromagnetisms}

Before we proceed, we define fractional derivatives as they are somewhat non-standard in theoretical physics though they have been used extensively to formulate fractional diffusion equations in the context of anomalous classical transport\cite{klafter}.  Although fractional derivatives date back to a letter between L'H${ \rm \hat o}$pital and Leibniz in 1635, it was not until 1832 that Liouville introduced a firm mathematical footing\cite{Millercalc,spanier}.  While it is standard to define fractional derivatives in terms of Fourier and Melin integral transforms\cite{Millercalc,spanier}, what is odd about fractional derivatives can be understood simply by replacing the factorial with the gamma function in the standard formula for differentiating the monomial, $x^k$,  to obtain,  
\beq
\frac{d^a}{dx^a}x^k=\frac{\Gamma(k+1)}{\Gamma(k-a+1)} x^{k-a}.
\eeq
As is evident, even for a constant ($k=0$), the fractional derivative is non-zero but vanishes when $a$ is an integer.  The utility of integer derivatives is that most functions can be well approximated in the region of interest by a line.  On fractals, as has been advocated recently for the underlying geometry in the pseudogap regime\cite{bianconi}, such is not the case and hence fractional derivatives of constants do not conform to the standard expectation.  Indeed, fractional electromagentisms have been  formulated\cite{lazo2011,hermann} previously on purely formal grounds, unlike the work of Pippard's\cite{pippardref}.  However, all such previous approaches are inspired by a simple generalization of Maxwell equations using the fractional derivative defined earlier rather than from the inherent extra degree of freedom that the zero-eigenvalue in Eq. (\ref{mgen}) entails.   In short, all such theories invoke the gauge transformation
\beq\label{fracinv}
A_\mu\rightarrow A_\mu+\partial_\mu^\alpha\Lambda,  
\eeq
where $\partial_\mu^\alpha$ is the fractional derivative. 

There are three main problems that arise from formulations based on Eq. (\ref{fracinv}).  The first is that the formulations based on fractional derivatives, due to the lack of a simple form of the chain rule, are dependent on a specific choice of coordinates, which makes it impossible to define a meanigful physical theory. The second and equally important,  rotational or Lorentz symmetry (depending on signature)  is absent as one easily sees from switching to momentum space.  The third is that $A_\mu$ no longer transforms as a a tensor, and {\it a fortiori} as a ~1-form.  Consequently, there are no gauge transformations that one can define. These problems are not just a feature of the classical theory as they in fact make it impossible to define any meaningful quantization.

\subsection{N\"other's Second Theorem}

N\"other's first theorem undergirds much of modern theoretical physics.  However, it is the second theorem which actually foreshadows fractional electromagnetisms.  The essence of the first theorem is that associated with any symmetry that can be formulated in terms infinitesimal variations of the action is a conserved current.  That is, differential symmetries of local actions imply conserved currents.  For the problem at hand, the relevant statement can be generated by the action 
\beq
S=\int d^dx[-\frac{1}{4}F^2+J_\mu A^\mu+\cdots].
\eeq
Since the field strength, $F$, is invariant under Eq. (\ref{eq:U1transformation}), the action transforms as
\beq
S\rightarrow S+\int d^dx J_\mu\partial^\mu\Lambda,
\eeq
under the gauge symmetry.
Consequently, invariance under Eq. (\ref{eq:U1transformation}), upon integration by parts, implies the standard 
charge conservation equation
\beq\label{chargecons}
\partial^\mu J_\mu=0.
\eeq
N\"other's second theorem arises, in essence, from an ambiguity in the first theorem.  Namely, Eq. (\ref{chargecons}) is unique up to any differential operator, which we call $\hat Y$, that commutes with the divergence, $[d,\hat Y]=0$.  Hence, additional constraints are needed to uniquely specify the current.  However, any such constraints will also appear in the generator of the gauge symmetry.  What N\"other noted in her original paper\cite{nother} is that by generalizing the gauge transformation
\beq
\label{Nothergen}
A_\mu\rightarrow A_\mu +\partial_\mu\Lambda +\partial_\mu\partial_\nu G^\nu+\cdots,
\eeq
to include arbitrarily high derivatives, additional constraint  equations can be derived for the current.  In fact, this procedure leads to a family of conserved currents of arbitrary rank.  Note, even the order of the form of the gauge field is now variable.  Until now, such high-order derivatives have not found any use in gauge theories because they generate no new information for the simple reason that if $\partial_\mu J^\mu=0$, then so do any higher-order integer derivatives.  In fact, in a recent paper\cite{avery}, it was highlighted that no physical consequences have been found thus far for such higher-order derivatives in the gauge expansion.

There is of course an overlooked possibility which yields non-trivial results.  The ``higher-order" derivatives in Eq.  (\ref{Nothergen}) need not be integer.  To obtain a rotationally invariant theory, the simplest possibility is the fractional Laplacian \begin{footnote}{More generally, one could consider an operator $L$ whose Fourier transform were $\hat L(f)= \varphi (|p|^2)\hat f$, for some function $\varphi(t)$, and $|p|^2$ is either in Euclidean or Lorentzian signature.  }\end{footnote}, namely the inclusion of fractional Laplacians in the generation of the gauge transformation. The importance of this observation is in the fact that indeed one can actually introduce lower order operators, using the fractional Laplacian. One then obtains conservation laws which are of the form
\beq \label{fractchargecons}
\partial^\mu (-\Delta)^{(\gamma-1)/2}J_\mu=0.
\eeq
\noindent
In such a theory, conservation laws such as the one in Eq. (\ref{fractchargecons}) are in some sense more fundamental, as one can infer the standard ones from them but they occur earlier in the hierarchy of conservation laws that stem from N\"other's first theorem.
This is the same conclusion reached from the degeneracy of the eigenvalue of Eq. (\ref{mgen}).   This consilience is not surprising because the degeneracy of the eigenvalue is another way of stating N\"other's second theorem.  That is, the current is not unique in gauge theories.

\section{Holographic Models with Fractional Gauge Transformations}

The precursors to this section show that electromagnetisms other than that governed by Eq. (\ref{eq:U1transformation}) are, in principle, possible without any essential feature of a gauge theory being altered.  As long as the underlying theory contains only local interactions, due to Gross' argument explained in the introduction (which one can think of as a sort of no-go theorem), we are stuck with the current formulation of electricity and magnetism.   A way out of this conundrum is to investigate higher-dimensional theories and construct the boundary theory explicitly.   A typical feature of holographic constructions is that bulk gauge fields simply act as sources\cite{Witten1998,klebanov} for global $U(1)$ currents at the boundary.  That is, the boundary, which we denote by the zero of the radial coordinate, $y=0$, is not imbued by a local gauge structure in which $A(y=0,x)=A_\parallel +d\Lambda$.  More explicitly, once the boundary condition is set, $A(y=0,x)=A_\parallel$, the gauge degree of freedom is lost. However, since the coupling at the boundary is simply $A_\parallel J$, any claim that holographic models yield non-trivial dimensions for the current at the boundary would require that $A$ also has an anomalous dimension and hence the boundary must then have a non-trivial gauge structure.  Large gauge transformations\cite{avery} must then come into play here since these are precisely the transformations that are non-vanishing for a boundary at infinity as in pure anti de-Sitter or Lifshitz spacetimes.

A result of bulk dilatonic holographic theories\cite{g1,Gouteraux,kiritsis} is that they give rise to boundary theories which have a non-traditional dimension for the gauge field and the associated $U(1)$ current.  Such an outcome requires a structure of the boundary theory beyond the standard procedure of dualizing the bulk gauge field as a source for the current at the boundary.  That is, it necessitates a non-trivial gauge structure at the boundary.  To show how this state of affairs obtains, we consider the precise form of the dilatonic action,
\beq 
S=\int \mathrm{d}^{d+1}xdy\sqrt{-g}\left[\mathcal R-\frac{\partial\phi^2}2-\frac{Z(\phi)}4F^2+V(\phi)\right],\nonumber\\
\eeq
that has been shown\cite{g1,Gouteraux} to give rise to anomalous dimensions for the boundary gauge field.  Asymptotically ($\phi\rightarrow\infty$), the dilaton field has the form,
 \beq
\lim_{\phi\rightarrow\infty}Z(\phi)\rightarrow Z_0 e^{\gamma\phi}\approx y^a,
\eeq
where $a\in {\cal R}$. Consequently, the Maxwell equations for this action reduce to
\beq\label{dilaton}
\nabla^\mu( y^aF_{\mu\nu})=0.
\eeq
In the language of differential forms, this equation becomes\begin{footnote}{ The $\star$ operator sends a ~p-form to an $n-p$-form and if $g_{\mu \nu}$ are the components of the metric and $\omega = \omega _{\alpha _1\cdot \alpha _p} dx_{\alpha _1} \wedge \cdots \wedge dx_{\alpha _p}$ is a ~p-form then by definition the components of $\star \omega$ are given by $(\star \omega) _{\mu_1 \cdot \mu_{n-p}}=  \frac{\sqrt{|g|}}{p\!} \epsilon _{\mu_1 \cdot \mu_{n} } g^{\mu _{n-p+1}\nu_1}\cdots  g^{\mu_n\nu_p}\, \omega_{\mu_1 \cdot \mu_{p}}$. }\end{footnote}
\beq\label{astar}
d(y^a\star dA)=0,
\eeq
which clearly illustrates that along any slice perpendicular to the radial direction, the standard $U(1)$ gauge transformation applies. At the boundary, the equations of motion yield no information and hence a more subtle method is needed to deduce the gauge structure at the boundary as the argument in the parenthesis vanishes at $y=0$.  However, it is evident from an inspection of the purely Maxwell part of the bulk action,
\beq
S_{\rm Max}=\int dV_d dy [y^a F^2+\cdots] ,
\eeq
that the dimension of the bulk gauge field is non-traditional: $[A]=1-a/2$.  This realization answers an obvious question: How can the gauge field acquire an anomalous dimension in holography since the radial coordinate amounts to renormalization, and it is well known that no amount of renormalization can change the engineering dimension of the gauge field\cite{gross,peskin,wen1992scaling}.  The dimension of $A$ is fixed to the engineering dimension it acquired in the bulk from the dilaton fields.  Since the dilaton vanishes at the boundary,
a direct derivation of the boundary action is necessary to make explicit the dimension of the gauge field.  {\it A priori}, we know the dimension must be non-traditional because the dilaton couples to the UV current.

\subsection{Caffarelli/Silvestre Extension Theorem}

The equations of motion for the gauge field, Eq. (\ref{astar}), are highly suggestive of an analysis along the lines of a well known theorem due to Caffarelli/Silvestre\cite{CS2007}.  In 2007, Caffarelli and Silvestre (CS)\cite{CS2007} proved that standard second-order elliptic differential equations in the upper half-plane in ${\mathbb R}_+^{n+1}$ reduce to one with the fractional Laplacian, $(-\Delta)^\gamma$, when one of the dimensions is eliminated to achieve ${\mathbb R}^n$.  For $\gamma=1/2$, the equation is non-degenerate and the well known reduction of the elliptic problem to that of Laplace's obtains. The precise statement of this highly influential theorem is as follows.  Let $f(x)$ be a smooth {\it bounded} function in ${\mathbb R}^n$ that we use to solve the extension problem
\beq
g(x,y=0) &= &f(x) \nonumber\\
\Ly g &=& 0 \label{eq:withy}
\eeq
to yield a smooth {\it bounded} function, $g(x,y)$ in ${\mathbb R}^{n+1}_+$.
In these equations $f(x)$ functions as the Dirichlet boundary condition of $g(x,y)$ at the boundary $y=0$.  These equations can be recast in degenerate elliptic form,
\beq
{\rm div}(y^a\nabla g)=0\quad{ \in}\,{\mathbb R}_+^{n+1},
\eeq
which CS proved has the property that 
\begin{equation}\label{caff-limit}
  \lim _{y\to 0^+} y^a \frac{\partial g }{\partial y} =C_{n,\gamma}\; {(-\lap)^\gamma f} 
\end{equation}
for some (explicit) constant $C_{n,\gamma}$ only depending on $d$ and $\gamma = \frac{1-a}{2}$
with $(-\lap)^\gamma$, the Reisz fractional Laplacian defined earlier.
That is, the fractional Laplacian serves as a Dirichlet to Neumann map for elliptic differential equations when the number of dimensions is reduced by one.  Consider a simple solution in which, $g(x,0)=b$, a constant, but also $g_x=0$.   This implies  that $g(y)=b+y^{1-a}h$ with $(1-a)>0$.  Imposing that the solution be bounded as $y\rightarrow\infty$ requires that $h=0$ leading to a vanishing of the LHS of Eq. (\ref{caff-limit}).  The RHS  also vanishes because $(-\Delta_x)^\gamma b=0$.  As a final note on the theorem, from the definition of the fractional Laplacian, it is clear that it is a non-local operator in the sense that it requires knowledge of the function everywhere in space for it to be computed at a single point.  In fact, it is explicitly an anti-local operator.  Anti locality of an operator $\hat T$ in a space $V(x)$ means that for any function $f(x)$, the only solution to $f(x)=0$ (for some $x\in V$) and $\hat T f(x)=0$ is $f(x)=0$ everywhere.  Fractional Laplacians naturally satisfy this property of anti-locality as can seen from their Fourier transform of Eq. (\ref{reisz}). 

\subsection{p-Form Generalization of Caffarelli/Silvestre}

To obtain the boundary theory in the case of the dilaton action, all that is necessary is a generalization of the Cafarelli/Silvestre theorem to a 1-form.  For the sake of generality,
we proved the theorem for any p-form\cite{csforms}.  Here we provide a simplification of our {\it Comm.  Math. Phys.}\cite{csforms} proof.  Essential to the proof is the new concept of the fractional differential rather than the usual fractional derivative.   Let us define $\Omega^p(M)$ as the space of p-forms on a manifold $M$.  The standard differential operator maps a p-form to a $p+1$-form while the adjoint differential operator, $d^\ast$ does just the opposite: $d^*:  \Omega ^p(M)\to  \Omega ^{p-1}(M)$.  Since the Hodge Laplacian does not change the order of a p-form, it is natural to define this operator,
\beq
\Delta=dd^\ast+d^\ast d:\Omega ^p (M)\to  \Omega ^p (M),
\eeq
in terms of products of $d$ and $d^\ast$. Following the spectral theorem, we defined \cite{csforms} the fractional Laplacian on forms as
\beq \label{fract-lap-def}\Delta ^\gamma \alpha= \frac{1}{\Gamma (-\gamma)} \int _0^{\infty} \; \left( e^{-t \Delta} \alpha - \alpha\right) \frac{dt}{t^{1+\gamma}},\eeq
for $\gamma\in (0,1)$ and for negative powers, we define
\beq\label{fractionalintegral}
\Delta ^{-s} \omega = \frac{1}{\Gamma(s)}\int_0^{+\infty} e^{-t\Delta} \omega \; \frac{dt}{t^{1-s}},
\eeq
with $s>0$.
As one does for the fractional Laplacian on functions, we define
\beq
\Delta ^\gamma=\Delta ^{\gamma -\lfloor{\gamma}\rfloor} \Delta ^{\lfloor{\gamma}\rfloor},
\eeq
where $\lfloor{\gamma}\rfloor$ indicates the integral part of $\gamma$.
In fact, this makes sense for any self-adjoint operator and in particular it applies to both $dd^*$ and $d^*d$.
Here, the heat {\it semigroup} $e^{-t \Delta} \alpha$ on forms is defined by requiring that $\beta=e^{-t \Delta} \alpha$ be the solution to the diffusion equation,
\beq\label{semigroup}
\frac{\partial} {\partial t}  \beta + \Delta \beta = 0,
\eeq
with initial condition $\beta (x,0) = \alpha (x)$.
One of the new technical novelties in \cite{csforms} is the definition of the fractional differential as
\beq
d_\gamma \omega = \frac{1}{2}\left( d\Delta ^\frac{\gamma-1}{2}\omega+ \Delta ^\frac{\gamma-1}{2}d \omega\right)
\eeq
which can in fact be rewritten as 
\beq \label{dgamma}
d_\gamma= d\Delta ^\frac{\gamma-1}{2}= \Delta ^\frac{\gamma-1}{2} d,
\eeq
since one readily shows that $[d, \Delta ^b]=0$ for any power $b$.
One of the important virtues of $d_\gamma$ is that it  behaves as the standard differential for the purpose of calculating 

\beq
\Delta^\gamma \omega=d_\gamma d_\gamma ^\ast+ d_\gamma ^\ast d_\gamma,
\eeq
the fractional Lapacian on any p-form $\omega$.  

With these definitions, we  proved\cite{csforms} that  for $\alpha\in\Omega^p$ and a bounded solution to the extension problem
\beq
\label{formsCS}
& d(y^ad^*\alpha)+d^*(y^a d\alpha)=0\in M\times \mathbb R_+\nonumber\\
& \alpha \mid _{\partial M} = \omega \text{ and } d^* \alpha \mid _{\partial M} =d_x^* \omega,
\eeq
then
\beq
\lim _{y\to 0} y^{a} {\rm i } _{\nu} d\alpha=C_{n,a} (\Delta) ^\gamma \omega,
\eeq
with $2\gamma= 1-a$ and where ${\rm i} _V \omega$ indicates the $(p-1)$-form determined by 
 ${\rm i} _V \omega (X_1,\cdots, X_{p-1})= \omega (X_1,\cdots, X_{p-1}, V)$,  $\nu = \frac{\partial}{\partial y}$, for some positive constant $C_{n,a}$.  This is the p-form generalization of the Caffarelli/Silvestre extension theorem.  It implies that the CS extension theorem on forms is the CS extension theorem on the components of the p-form.  The succinct statement in terms of the components is easiest to formulate from the equations of motion
 \beq\label{formsCS2}
& {\rm div} (y^a \nabla \alpha _{i_1\cdots i_p})=0\in M\times \mathbb R_+\nonumber\\
& \left( \alpha _{i_1\cdots i_p}\right)\ \mid _{\partial M} = \omega _{i_1\cdots i_p}\ \text{ and } d^* \alpha \mid _{\partial M} =d_x^*\omega.
\eeq
Therefore, using the CS theorem, we have that 
\beq
\lim_{y\rightarrow 0} \; y^a \frac{\partial \alpha _{i_1\cdots i_p}} {\partial y} =C_{n,a}(-\Delta)^\a \ \omega _{i_1\cdots i_p},
\eeq
which proves that
\beq
\lim _{y\to 0} y^{a} {\rm i } _{\nu} d\alpha=(\Delta) ^a \omega,
\eeq
since by (elliptic) regularity of solutions to Eq. (\ref{formsCS})
\beq
\lim _{y\to 0} y^{a} \frac{\partial \alpha _{0 \ell _1, \cdots \ell _{p-1}} }{\partial x^{j_k}} =0.
\eeq

\begin{figure}
	\includegraphics[scale=0.3]{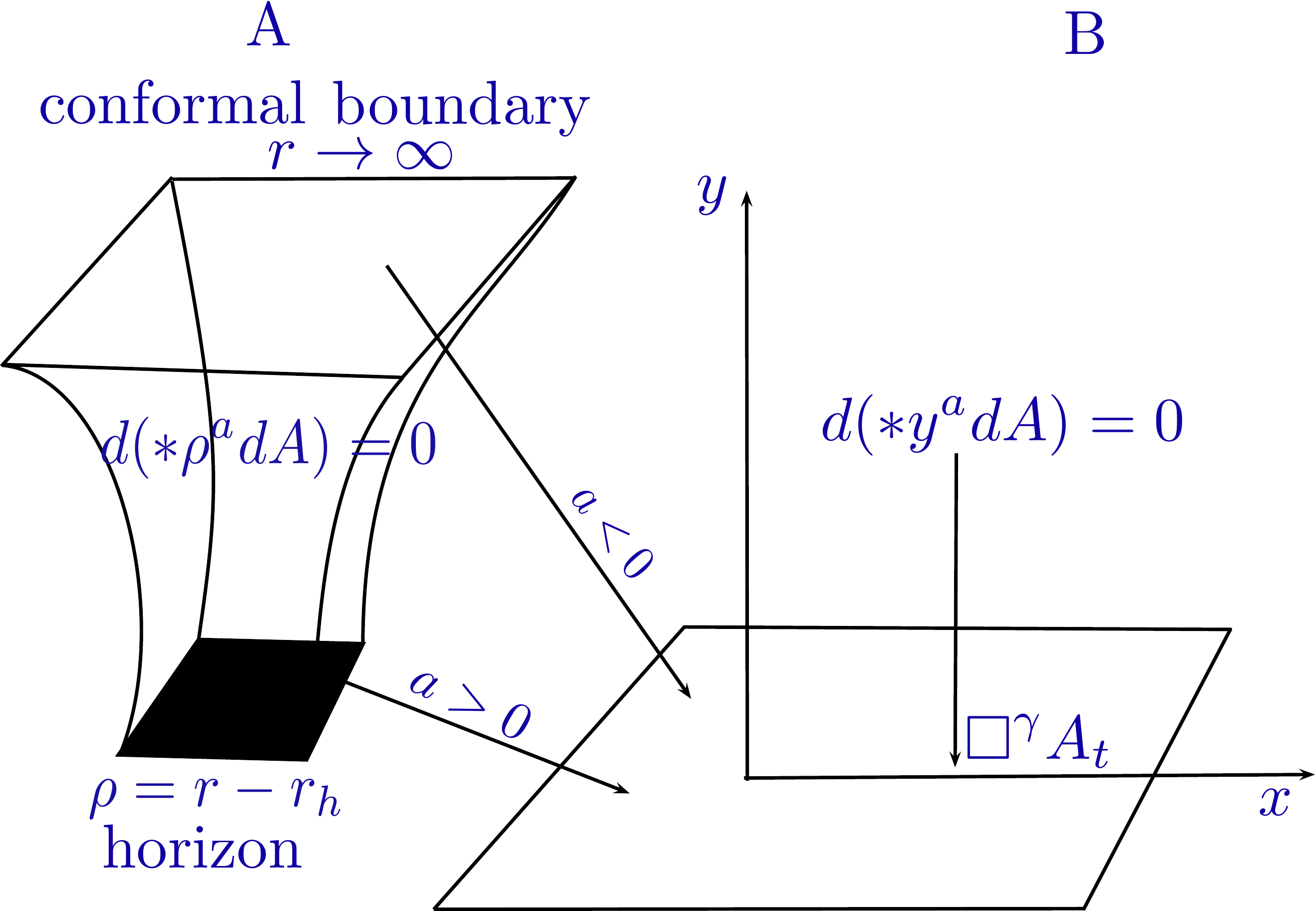} 
    \caption{A.) A depiction on an AdS spacetime with a conformal boundary at $r=\infty$ and a black hole horizon at $r=r_h$.  The Maxwell-dilaton action in the bulk has equations of motion of the form  $d(\ast \rho^a dA)=0$.  B.)  p-form generalization of the Caffarelli-Silvestre\cite{CS2007}-extension theorem. $A_t$ are the boundary (tangential) components of the bulk gauge field, $A$. For a dilaton action in $\mathcal R^n$ with the equations of motion $d(\ast y^a dA)=0$, the restriction of these equations of motion to the boundary yields the fractional Box operator where the exponent is given by $\gamma=(1-a)/2$.  Depending on the  sign of $a$, the bulk dilaton action either yields fractional Maxwell equations of motion at the conformal UV $(a<0)$ boundary or at the IR limit $(a>0)$ demarcated by the horizon radius, $r_h$.  } \label{pform}
\end{figure}

Applied to the dilaton action in Eq. 
(\ref{dilaton}) we see that the boundary theory obeys a fractional Maxwell equation of the form
\beq\label{fracmax}
\Delta^\gamma A_t=0.
\eeq
This theorem is equally valid at the black hole horizon, the IR limit, using the membrane paradigm\cite{thorne}.  Figure (\ref{pform}) depicts the generality of the theorem proved here.  As shown, depending on the sign of $a$, the CS extension theorem applied to p-forms either yields the fractional Maxwell equations at the UV conformal boundary, $a<0$, or at the horizon (IR limit), $a>0$.  Hence, the theorem is completely general and does is not restricted to the UV boundary.

 The curvature that generates these boundary equations of motion is 
\beq
F_\gamma = d_\gamma A= d\Delta ^\frac{\gamma-1}{2} A,
\eeq
with gauge-invariant condition,
\beq\label{eq:u1frac}
A \rightarrow A+ d_\gamma \Lambda,
\eeq
 where the fractional differential is as before in Eq. (\ref{dgamma}) which preserves the 1-form nature of the gauge field.
 This feature is guaranteed because by construction, the fractional Lagrangian cannot change the order of a form.
 As is evident,
 $[A_\mu]=\gamma$, rather than unity.  This gauge transformation is precisely of the form permitted by the preliminary considerations on N\"other's second
 theorem presented at the outset of this article and also consistent with the zero eigenvalue of the matrix $M$ in Eq. (\ref{Maction}).
 
There are two alternative ways to obtain the fractional Maxwell equations derived earlier.  As shown by Domokos and Gabadadze\cite{domokos}, one can start with a bulk theory with a field strength  and gauge transformation of the form 
\beq
F_{\mu\nu}=\partial_\mu A_\nu-\partial_\nu A_\mu,\quad A_\mu\rightarrow A_\mu+\partial_\mu\Lambda\nonumber\\
F_{\mu y}=\partial_\mu A_y-\partial_y^\alpha A_\mu,\quad A_\mu\rightarrow A_y+\partial^\alpha_y\Lambda
\eeq
and then integrate out the radial ($y-$) degrees of freedom.  The result is a boundary theory with fractional Maxwell equations of motion identical to those that arise from the CS extension theorem on p-forms derived earlier.  While this mechanism is not as general as the bulk dilaton coupling, the key in this derivation is that a fractional gauge transformation along the radial direction only is sufficient to yield a boundary theory with fractional Maxwell equations.   In the spirit of this derivation, one can introduce a term in the bulk action with a mass for the gauge field along the radial direction only: $m^2 A_y^2$.   Simply apply the CS theorem as found earlier\cite{lpprd} to the bulk action and immediately the boundary theory obtains with fractional Maxwell equations identical to those in Eq. (\ref{fracmax}).  The dimension of the current at the boundary or the UV is determined by the mass of the bulk radial gauge field. What all of these derivations imply is that bulk IR degrees of freedom that  either change the dimension of the gauge field in the bulk or give the gauge field a mass but only along the radial direction dictate the dimension of an effective IR operator which overlaps with the UV current. Since both of these schemes result in the same boundary theory, it is the breaking of $U(1)$ symmetry in the higher dimensional manifold that gives rise to the non-local electromagnetism at the boundary and hence this resonates with the Pippard kernel in superconductivity. Quite generally, the origin of novel scaling for currents and gauge fields is the breaking of $U(1)$ symmetry to ${\mathbb Z}_2$ either explicitly (BCS superconductivity) or implicitly as in the boundary action resulting from bulk dilaton theories.  In both cases, non-local electromagnetism obtains.  Our treatment then puts Pippard's analysis of the Meissner effect in a new light.  Namely, the non-locality arises because the associated current has an anomalous dimension just as in the holographic dilatonic models.  

\subsection{Causality}

The theory that one infers from a non-local Lagrangian such as the fractional Maxwell theory,
\beq\label{fractionalLag}
S = \int dV_d(-\frac{1}{4}|d_\gamma A|^2+ J^\mu A_\mu),
\eeq
is still causal. One can see this both classically and quantum mechanically. The classical argument boils down to showing that plane waves which are solutions of the vacuum fractional equations, travel at the speed of light. 
Indeed, the fractional Maxwell equations can be written, in the language of differential forms, as 
\beq
d_\gamma F_\gamma =0 , \qquad d_\gamma ^* F_\gamma=J,
\eeq
where $J$ is the current. In the vacuum (i.e. $J=0$) and in a gauge in which $d^* A=0$ (i.e., $\partial _\mu A^\mu=0$), these equations become,
\beq\label{boxA}
\Box ^\gamma A^\mu=0,
\eeq
whence one derives 
\beq\label{boxF}
\Box ^\gamma F^{\mu\nu}=0.
\eeq
Here, $\Box = -\Delta +\frac{\partial^2}{\partial t^2}$ is the Box operator and we are assuming throughout the section that the speed of light is $c=1$.
We next note that any solution to such an equation can be written as a superposition of waves $e^{i\left(\vec{k}\cdot \vec{x}-\omega t\right)}$ (via Fourier transform) and observe that $e^{i\left(\vec{k}\cdot \vec{x}-\omega t\right)}$ is an eigenvalue of the box operator $\Box$, since
\beq
\Box e^{i\left(\vec{k}\cdot \vec{x}-\omega t\right)}= (k^2- \omega ^2) e^{i\left(\vec{k}\cdot \vec{x}-\omega t\right)}.
\eeq
As  $[\Box^\gamma, \Box]=0$, the two operators share the same complete basis of eigenfunctions and therefore
\beq
\Box ^\gamma \left(e^{i\left(\vec{k}\cdot \vec{x}-\omega t\right)}\right)= (k^2- \omega ^2)^\gamma e^{i\left(\vec{k}\cdot \vec{x}-\omega t\right)},
\eeq
which entails that solutions to $\Box ^\gamma F^{\mu\nu}=0$ must be (superpositions of) waves traveling at the speed of light, since $k^2=\omega ^2$.
From a quantum field theory stand point, propagators will not be zero outside of the light-cone (and indeed this holds even in the local theory). In terms of a direct proof from the boundary operators, one can show, using the Caffarelli-Silvestre extension theorem and its generalizations given any field $\phi$ in the theory, one has that
\beq
[\phi(x), \phi(y)]=0,
\eeq
provided $(x-y)^2=0$.  Hence, there is no problem with causality though the theory is non-local.

\subsection{Pippard Reloaded}

As remarked earlier, the appearance of the Pippard kernel in the current, Eq. (\ref{pipJ}), imbues the current with a non-traditional or ``anomalous'' dimension.  Consequently,
it should be possible to recast the Pippard kernel within the fractional formalism we have derived here.  To see how this comes about, we consider the 
complex Klein-Gordon Lagrangian,
\beq
L= \partial _\mu\psi \partial ^\mu \psi^*- m^2 \psi \psi^*,
\eeq
for the matter content of a BCS superconductor.
As is evident from N\"other's first theorem, the conserved current, given by 
\beq
J^\mu= i\left(\psi*\partial ^\mu \psi- \psi \partial ^\mu \psi^*\right),
\eeq
 does not have the correct scaling dimension, or ``anomalous" dimension, (necessary from Gross's argument) to describe anomalous transport consistent with the Pippard kernel.   Some other ingredient is required as far as the matter content of the Lagrangian is concerned.  The remedy is to interject  a mechanism to introduce a current with scaling dimension, of the form \beq J^\mu= i\left(\phi*\partial ^\mu \Box ^{(1-\gamma)/2}\phi- \phi\partial ^\mu \Box ^{(1-\gamma)/2}\phi^*\right)\eeq 
into the Lagrangian rather than the standard linear derivatives
This has the effect of changing the Lagrangian to 
\beq
L= \partial _\mu\Box ^{(1-\gamma)/2}\phi\partial ^\mu \Box ^{(1-\gamma)/2}\phi^*- m^2 \phi \phi^*,
\eeq
This is achieved by imposing that the gauge transformation on $A_\mu$ is fractional,
\beq\label{gaugeonA}
A_\mu \rightarrow A_\mu + \partial_\mu\Box^{\frac{\gamma-1}{2}}\Lambda\equiv A^\prime_\mu,
\eeq
for some dimensionless function $\Lambda$. The corresponding covariant derivative is $D_{\gamma, A}\phi=\left( \partial _\mu + i e \Box ^{(1-\gamma)/2} A^\mu\right) \Box ^{(1-\gamma)/2}  \phi$.  The action of the gauge group is taken to be,
\beq 
e^{i\Lambda } \odot \phi=  \Box ^{(\gamma-1)/2} \left(   e^{i \Box ^{(1-\gamma)/2} \Lambda }  \phi\right)
\eeq
so that one has, in a fashion reminiscent of standard $U(1)$ gauge theory
\beq
D_{\gamma, A}(e^\Lambda \odot \phi)=e^{i \Box ^{(1-\gamma)/2} \Lambda } D_{\gamma, A'}\phi,
\eeq
where $A^\prime_\mu= A_\mu $ is defined implicitly from the action induced by the gauge transformation, as in eq. \eqref{gaugeonA}. This is consistent with standard transformations of Gauge, if one performs the non-local transformations 
$\Phi =  \Box ^{(1-\gamma)/2}  \phi$, $a=  \Box ^{(1-\gamma)/2}  A$ and $\lambda =  \Box ^{(1-\gamma)/2}  \lambda$. Note that this field redefinitions do not give rise to a standard theory of matter plus Gauge field, as the quadratic term in $\phi$ in the Lagrangian does not transform to a quadratic term in $\Phi$.

One can immediately write down the correct gauge invariant (emergent) theory as
\beq
L=D_{\gamma, A}\phi (D_{\gamma, A}\phi)^*- m^2 \phi^*\phi- F^{\mu \nu} _\gamma {F_{\mu \nu}}_\gamma,
\eeq
where $ F^{\mu \nu} _\gamma = \partial _\mu \Box ^{(\gamma-1)/2} A_\nu - \partial _\nu \Box ^{(\gamma-1)/2} A_\mu$ and can be interpreted as the commutator $[D_A, D_{\gamma, A}]$.
Simply by choosing $\gamma$ appropriately introduces a current consistent with the dimension $d-d_C-1$ of the Pippard kernel.  Hence, both the Meissner effect and holographic dilatonic models can be described by the same formalism.  
For the non-abelian case, there are various possibilities. The most natural one would be to take, as normally done for local operators, the curvature to be the commutator of the covariant derivative. The problem is that $D_{A,\gamma}\phi = d\Box ^{(\gamma-1)/2} \phi+ i g \Box ^{(\gamma-1)/2} A,\Box ^{(\gamma-1)/2} \phi$ is not a derivation. We define the curvature via
\beq [D_{\Box ^{(\gamma-1)/2} A} , D_{\gamma, A}]= i F\eeq 
so that $F_{\mu\nu}=  \partial _\mu \Box ^{(\gamma-1)/2} A_\nu - \partial _\nu \Box ^{(\gamma-1)/2} A_\mu+ i g [\Box ^{(\gamma-1)/2} A_\mu, \Box ^{(\gamma-1)/2} A_\nu]$, thereby completing the gauge structure of this theory.

\subsection{Fractional Virasoro Algebra}

Does a Virasoro algebra control the algebraic structure of currents that have anomalous dimensions?  In string theory, the Virasoro algebra underpins the conformal structure of the local current operators.  The Virasoro algebra is the central extension of the Witt algebra on the space of {\it local} conformal transformations on the unit disk.  In fact,  for any problem controlled by critical scaling, the Virasoro algebra governing the conserved currents is of fundamental importance. In all constructions of the Virasoro algebra thus far, the generators are entirely local.  The standard generators of the Witt algebra are the standard angular momentum operators
\beq
L_n :-z^{n+1}\frac{\partial}{\partial z},
\eeq
with $z=\xi_1+i\xi_2$.
These operators obey a simple commutation relation $[L_n,L_m]=(n-m)L_{n+m}$ forming the Witt algebra and a central extension that can be established from the Jacobi identity of the form,
\beq
[L_n,L_m]=(n-m)L_{n+m}+\frac{c}{12}m(m^2-1)\delta_{m+n,0}.
\eeq
 
 A similar but more complicated structure governs the algebraic features of fractional currents relevant to the boundary theories introduced here.  To disclose this structure,
 we consider the generalization
\begin{equation}
L^a_n= -z^{a(n+1)} \pa ,\qquad {\bar L } ^a_n:=-\bar z^{a(n+1)} \pab
\end{equation}
acting on $V^a:=\CC[ [z^{-a}, z^a]]$. In this algebraic description we think of $z^a$ merely as a formal expression as in Puiseaux series, if $a\in \mathbb Q$ \cite{e2000}.
The algebra $\mathcal W_a$ has a special structure, which we called a Lie multimodule; namely, there are operations $\star _{(p,q)}$ on $\mathcal H$ and a grading on $\mathcal W_a$, such that: $[\phi\otimes L_p, \psi \otimes L_q]= \phi \star _{p,q} \psi[L_p,L_q]$. 
All the central extensions which preserve this extra structure
\beq
0\to \mathcal H \to \mathcal V _a \to \mathcal W_a\to 0,
\eeq
which are parametrized by a group $H^2_\star(\mathcal W_a,\mathcal H)$ (which is isomorphic to $\mathcal H$) are of the form 
\beq
[L^a_m,L_n^a]=A_{m,n}L^a_{m+n}+\delta_{m,n}h(n)cZ^a
\eeq
where $c$ is the central charge ($c\in \mathcal H$), $Z^a$ is in the center of the algebra and $h(n)$ obeys the recursion relation,
\beq
 &h(2)=c \nonumber\\
&\frac{A_{-1, -m} \Gamma _{(-(m+1))}- A_{m,1} \Gamma _{m+1}}{A_{-(m+1), m+1}}  \,  \,h((m+1)) \\=& \frac{A_{m+1, -1} \Gamma _1 
-A_{1,-(m+1)} \Gamma _{-m}}{A_{m,-m}}  \,h(m).
\eeq

Here
\beq A_{p,q}(s)=  \frac{\Gamma (a(s+p)+1)}{\Gamma (a(s-1+p)+1)} - \frac{\Gamma (a(s+q)+1)}{\Gamma (a(s-1+q)+1)}\nonumber\\
  \eeq
and 
\beq \Gamma _p(s)= \frac{\Gamma (a(s+p)+1)}{\Gamma (a(s-1+p)+1)}\eeq
where $\Gamma$ is the gamma function. 
The elements of $\mathcal W_a$ are operators acting on $\mathbb C [[z^{a}, z^{-a}]]$ via the prescription
\beq \left(\phi \otimes L_p^a\right) (z^{ka})= \phi(k)L_p^a (z^{ka}).\eeq

\noindent
The usual $\mathcal H$-Lie algebras $\mathcal V _a$ are a generalization of the Virasoro algebra in that
\beq
\lim _{a\to 1} \mathcal V_a= V.
\eeq

The Lie algebra structure of $\mathcal V_a$, on the other hand,  does not arise, for $a\neq 1$, as a tensor product of a Lie algebra $V$ with $\mathcal H$, reflecting the very non-local nature of the operators in $\mathcal V_a$. In this sense, it is a twisted structure, or more properly a Lie multimodule, further indicating the non local nature of non-local conformal field theories.

 \section{Application: Strange Metal}
 
 An immediate application of fractional electromagnetism is the normal state of the cuprates, dubbed the `strange metal,' a state of matter 
 which  is known to exhibit numerous power laws.  For example, 1) the resistivity is a linear function of temperature, 2) the Hall angle is a quadratic\cite{hallangle} function of temperature, 3) the Lorentz\cite{lorentz} number is not a number as would be the case for metals that obey the Weidemann-Franz law but in fact scales also as a linear function of temperature, and 4) the mid-infrared optical conductivity is proportional\cite{Marel2003,Hwang2007,Basov2011} to $\omega^{-2/3}$ for $\omega\gg T$.  The collection of these facts remains unresolved because no knock-down experiment has revealed unambiguously the nature of the charge carriers in the normal state.  While critical points imply power laws, the argument has been run in reverse in an attempt to resolve this problem.  Namely, the leading candidate\cite{qcrit2,qcrit3,varma} to explain the power laws is some type of quantum critical phenomenon.  In the simplest instance of quantum criticality, a single parameter governs all divergences.  Herein lies the inherent difficulty of this problem.  In the case of single parameter scaling, the temperature dependence of the resistivity is governed\cite{chamon} by $T^{(2-d)/z}$, whereas the frequency dependence\cite{wen1992scaling} of the conductivity scales as $\sigma(\omega)\propto\omega^{(d-2)/z}$.  As these exponents are negatives of one another, it is impossible within single-parameter scaling to explain the origin of T-linear resistivity and the mid-infrared scaling simultaneously.  In fact, the situation is much worse.  To obtain the correct temperature dependence of the resistivity one has to invoke either $d=1$, which is unphysical, or $z<0$ to explain just $T-$linear resistivity.  The latter violates causality.  
 
It has recently been suggested\cite{hk} on purely phenomenological grounds that the dc properties of the strange metal can be explained if the strange metal interacts with light such that the gauge field has an `anomalous' dimension such that
\beq
\left[A_i\right]&=&1-\Phi \\
\left[E\right]&=&1+z-\Phi \\
\left[B\right]&=&2-\Phi,
\eeq
with $\Phi=-2/3$.  As a consequence, $[A_i]=5/3$, a significant deviation from unity.  Both $E$ and $B$ contain anomalous dimensions as both are generated from the anomalous gauge field.  In standard Maxwell theory in which $[B]=2$, the magnetic flux through a tube of radius $r$ is simply $\pi r^2 B$.  This quantity is dimensionless as the area and the field have cancelling dimensions.  However, in the case of non-traditional scaling for the gauge field, $\pi r^2 B$ is no longer dimensionless, and hence this quantity cannot be the flux.  Precisely what is the flux will be resolved as will be the experimental implications of $[A_i]\ne 1$. 

\subsection{Mid-infrared Conductivity}

What about the mid-infrared scaling?  Experimentally, the AC conductivity in the strange metal phase\cite{Marel2003,Hwang2007,Basov2011} scales as $\omega^{-2/3}$ in the mid-infrared frequency range ($\sim$ 500 cm$^{-1}$ to 10000 cm$^{-1}$). This scaling is peculiar because it does not extend to $\omega=0$. At low frequencies, the AC conductivity obeys the Drude formula,
\beq
\sigma(\omega) = \frac{n e^2 \tau}{m} \frac{1}{1-i\omega \tau},
\eeq
with $n$ the charge density, $e$ the electric charge, $m$ the mass, and $\tau$ the relaxation time. 
However, since both the real and imaginary components of the conductivity both scale in this fashion\cite{Marel2003,Hwang2007,Basov2011}, explaining this  non-trivial result could involve physics not in addition to what is invoked to explain the dc transport properties.  What we show here is that the conductivity that scales in this fashion
offers excellent evidence for an anomalous dimension of the current.  What we invoked\cite{midir1,midir2} to explain the mid-infrared scaling is a multiscale sector in which all the system parameters run as a function of energy or equivalently mass:
\beq
n(m) &=& n_0\frac{m^{a-1}}{M^a} \\
e(m) &=& e_0\frac{m^b}{M^b} \\
\tau(m) &=& \tau_0\frac{m^c}{M^c},
\eeq
such that the conductivity is represented by the weighted sum
\beq \label{eq:con_weighted_sum}
\sigma(\omega) = \int_0^M \frac{n(m) e^2(m)}{m} \frac{1}{1-i\omega \tau(m)}dm.
\eeq
Here $n_0$, $e_0$, and $\tau_0$ are constants with the same units as density, charge, and relaxation time, respectively. The mass cutoff $M$ is an energy scale of the system, for example, the bandwidth. In terms of fundamental quantities, the exponents $a$ represent the hyperscaling violation exponent, $b$ the running of the charge and 
hence the anomalous dimension for the gauge field and $c$ governs the momentum relaxation.  
Such dependence on energy or mass has been invoked\cite{cm1,Georgi2007a,phillips2013,Karch2015} previously to model scale-invariant (dubbed unparticles) sectors.  Substituting all the mass distributions into Eq. (\ref{eq:con_weighted_sum}) results in the final expression 
\beq
\sigma(\omega) =  \frac{\rho_0e^2_0}{cM}\frac{1}{\omega(\omega\tau_0)^{\frac{a+2b-1}{c}}}\int\limits_{0}^{\omega\tau_0}dx \ \frac{x^{\frac{a+2b-1}{c}}}{1-ix},
\eeq
for the conductivity, where we have changed integration variables to $x = \omega\tau_0\frac{m^c}{M^c}$.  Define $\eta\equiv\frac{a+2b-1}{c}+1$. When $\eta<0$, the integral does not converge. This means it is {\it not} possible to obtain a positive power law in the optical conductivity from this model.  To reproduce the experiments we set $\frac{a+2b-1}{c} = -\frac{1}{3}$, so $\eta = \frac{2}{3}$.  We see then that to explain the experiments, the anomalous dimension need not be non-zero.  Taking the limit of $\tau_0\omega \rightarrow \infty$, we obtain
\beq
\sigma(\omega) = \frac{1}{3c}(\sqrt{3}+3i)\pi\frac{\rho_0 e^2_0 \tau^{\frac{1}{3}}_0}{M\omega^\frac{2}{3}},
\eeq
which exhibits the desired power-law scaling and has a phase angle of $60^\circ$ as is seen experimentally\cite{Marel2003,Hwang2007,Basov2011}.  While this procedure does not fix the value of $b$, consistency within the dc properties does. Applying the same mass summation to a free energy density, one finds that, in order to explain the scaling of the Lorentz ratio, $b\propto \Phi$ \cite{midir2}. Since $\Phi = -\frac{2}{3} \neq 0$, this means $b\ne 0$. Furthermore, by fitting with all the anomalous exponents obtained from the $dc$ properties, one finds $b = -\frac{1}{2}$ \cite{midir2}. As a result, a non-zero charge exponent ($b \neq 0$) leads to a non-trivial dimension for the gauge field ($\Phi \neq 0$ or, equivalently, $[A_i] \neq 1$) and also contributes to the mid-infrared scaling in the optical conductivity.  Because
$b$ and $\Phi$ are not reciprocally related, the quantity $qA$ does acquire
a non-trivial scaling dimension different from unity and hence is an
example of fractional electro-magnetism.

\subsection{Skin effect}

\begin{figure}
	\includegraphics[scale=0.5]{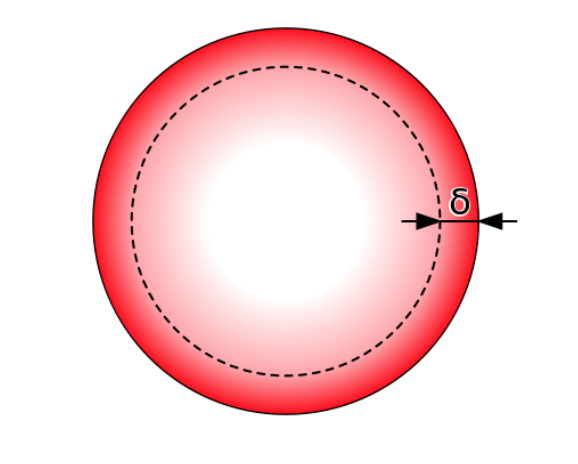} 
    \caption{In a conductor, the AC conductivity is confined to a narrow ribbon around the sample where the charge density accumulates, denoted the skin depth.  In a typical metal, this quantity scales as the square root with the resistivity because of the quadratic nature of the Maxwell equations.  The square root is modified in the fractional theory as in Eq. (\ref{newskin}). }  \label{skdepth}
\end{figure}

In a typical conductor, the AC conductivity is confined to a region (termed the skin depth) near the surface where the charge density accumulates as illustrated in Fig. (\ref{skdepth}). Because of the quadratic nature of the Maxwell equations, the skin depth is related to the resistivity, $\rho$, through
\beq\label{eq:delta_skindepth}
\delta=\sqrt{\frac{2\rho}{\omega\mu}},
\eeq
where $\mu$ is the permitivity of the medium and $\omega$ is the frequency.  In the strange metal, this expression will be modified because the Maxwell equations are no longer strictly quadratic.   Deriving the new skin effect is straightforward.  In component form, the Maxwell equations take the form\footnote{$\rho$ in Eq. (\ref{eq:gauss}) is the charge density, not to be confused with the resistivity in Eq. (\ref{eq:delta_skindepth}).},
\beq
\Box^{\frac{\gamma-1}{2}}\left(\nabla\times \vec{B} - \frac{1}{v^2} \frac{\partial \vec{E}}{\partial t}\right) &=& \mu \vec J \label{eq:ampere-maxwell} \\ 
\Box^{\frac{\gamma-1}{2}}\nabla \cdot \vec E &=& \frac{\rho}{\epsilon} \label{eq:gauss} \\
\Box^{\frac{\gamma-1}{2}}\left(\nabla\times \vec{E} + \frac{\partial \vec{B}}{\partial t}\right) &=& 0 \label{eq:faraday}\\ 
\Box^{\frac{\gamma-1}{2}}\nabla \cdot \vec B &=& 0 . 
\eeq
where $\Box = -\Delta +\frac{1}{v^2}\frac{\partial^2}{\partial t^2}$. Here $\mu$ is the permeability of the fractional medium and hence has non-traditional dimensions.  As is evident, the permittivity, $\epsilon$, also has non-traditional units. It is important to note that we take the wave speed and the quantity with units of speed in $\Box$ to be the same, i.e, $v = \frac{1}{\sqrt{\mu\epsilon}}$. 

For this fractional Maxwell equations it is not clear what are the conditions for a good conductor (so one can set the charge density $\rho = 0$). Nonetheless, we will assume that $\rho = 0$ (either because the system is neutral or the electric field has only the transverse component). From Eqs. (\ref{eq:ampere-maxwell}), (\ref{eq:gauss}), and (\ref{eq:faraday}), it follows that
\beq
\Box^{\frac{\gamma+1}{2}} \vec E  = -\mu \frac{\partial}{\partial t}\vec J.
\eeq
Using equation, $\vec J = \sigma \vec E$, the wave equation in a conductor is
\beq
\Box^{\frac{\gamma+1}{2}} \vec E  = -\mu\sigma \frac{\partial}{\partial t}\vec E.
\eeq
Suppose the sample occupies half the space ($x > 0$), and the wave propagates along the $+\hat x$ direction.  Using a plane wave solution,
\beq
\vec E = \vec E_0 e^{i(k x - \omega t)},
\eeq
we find that $k$ and $\omega$ must satisfy 
\beq
\left(k^2 -\frac{\omega^2}{v^2}\right)^{\frac{\gamma+1}{2}} = \frac{i\omega\sigma}{\epsilon v^2}.
\eeq
or
\beq
k^2 = \frac{\omega^2}{v^2} + \left(\frac{i\omega\sigma}{\epsilon v^2}\right)^{\frac{2}{\gamma+1}}.
\eeq
In the low frequency limit, $|\omega| \ll (\frac{\sigma}{\epsilon})^\frac{1}{\gamma}v^{\frac{\gamma - 1}{\gamma}} $, the second term dominates for $\gamma >0$,
\beq
k^2 = \left(\frac{i\omega\sigma}{\epsilon v^2}\right)^{\frac{2}{\gamma+1}}.
\eeq
One finds
\beq
k &=& k_1 + k_2 \nonumber \\
&=& \left(\frac{\omega\sigma}{\epsilon v^2}\right)^{\frac{1}{\gamma+1}}e^{i\left(\frac{\pi}{2(\gamma + 1)}+\frac{2\pi n}{\gamma + 1}\right)}, \nonumber \\
\eeq
where $n$ is an integer. Consequently, the skin depth is
\beq\label{newskin}
\delta = 1/k_2 = \left(\frac{\epsilon v^2}{\omega \sigma}\right)^{\frac{1}{\gamma + 1}}\frac{1}{\sin\left( \frac{\pi}{2(\gamma + 1)} + \frac{2\pi n}{\gamma + 1} \right)} .
\eeq
This power-law deviation from the standard square-root behaviour is a prediction that can be directly tested experimentally from AC measurements.  

\subsection{Aharonov-Bohm Effect and Quantization of Charge}

One of the basic constructions of ``anomalous" EM is that it is based on a new form $a \equiv \Box^{\frac{1-\gamma}{2}}A$ which is {\it classical}, in that, it satisfies the standard Maxwell equations. To illuminate this construct, consider the equations of motion,
\beq
d\Box^{\frac{\gamma-1}{2}}(\star d\Box^{\frac{\gamma-1}{2}}A)=\star J,
\eeq
 for the current in the case of non-traditional scaling for the gauge field. Taking just the spatial part of the $\Box$ operator to define the B-field, we have that
 \beq
 \int_\Sigma d_\gamma A=\oint_{\partial\Sigma} \tilde{A},
 \eeq
 with $\tilde{A}=\Delta^{\frac{(\gamma-1)}{2}} A$. 
 Although the equality follows from Stokes' theorem, the result does not seem to have the units to be a {\it quantizable} flux.  That is, it is not simply an integer $\times hc/e$.  The implication is then that the charge depends on the scale. We make this intuition precise in what follows, but first let us remark that the form $a$ satisfies the standard Maxwell equations.  
In fact, because $[d,\Box^\gamma]=0$, the equations of motion can be rewritten as
 \beq
 \Box^{\frac{\gamma-1}{2}}d(\star d\Box^{\frac{\gamma-1}{2}}A)=\star J.
 \eeq
 When these equations are invertible, they imply the existence of a new current and a new form $\tilde{A} \equiv \Box^{\frac{\gamma-1}{2}}A$,
 \beq \label{Maxtilde}
 d(\star d \tilde{A})=\star \Box^{\frac{1-\gamma}{2}} J\equiv \star j.
 \eeq
Consequently, the spatial part of the current $j$ is dual to $\tilde A$. We emphasize here that $\tilde A$ satisfies equations that are Maxwell-like.

Due to Weinberg's\cite{weinberg} argument in connection with Pippard's kernel in superconductivity, the current $J$ is a non-local expression of a field $A$. This is the field that appears in the path integral (including the fractional EM theory associated to $A$ ) and we therefore require, for the purposes of the well-posedness of the path integral, that $\int _\ell e A$ be an integral multiple of $h$.
 In order to express how $A$ interacts with matter, there are two possible alternatives. One is to transform $A$ to a new field $a$ in a manner that the ``local" gauge group acting induces the standard action $a\to a+d\Lambda$. Since the ``local'' gauge group acts on $A$ as $A\rightarrow A+d_\gamma \Lambda$, it is straightforward that, if we define $a= \Box ^{\frac{1-\gamma}{2}} A$, on $a$ it acts as
 $a_\mu \rightarrow a_\mu+\partial_\mu\Lambda$ and our physical theory is gauge invariant although $a_\mu$ is not directly related to the physical fractional gauge field. Also, since $a=\Box ^{1-\gamma}\tilde A$, from equation (\ref{Maxtilde}) it follows that $a$ satisfies the Maxwell equations
 \beq \label{Max-a}  d(\star d a)= \Box ^{1-\gamma} j= \Box ^{ \frac{3}{2}(1-\gamma)}J.\eeq
 so the current associated to $a$ is $\mathfrak J = \Box ^{ \frac{3}{2}(1-\gamma)}J$.

At this stage it would seem that nature is now forced to arbitrarily choose between $A$ and $a$ and that, all things being equal, the diligent practitioner should then choose $a$ as the correct physical object, following a mere criterion of simplicity as a deciding factor. Our observation is that in fact nature chooses heavily between making us face an alternative: either $a$ or $A$ is quantizable, not both.  This is the inherent physical consequence introduced by N\"other's Second Theorem: ambiguity in the gauge transformation leads to a breakdown of charge quantization.  

 In order to elucidate this alternative, we show in the Appendix that for every closed loop $\ell$,
 \beq \label{alternative}
{\mathrm Norm}\left( \int_\ell \tilde A\right)=\frac {\int_\ell A}{\Gamma(s+1)},
 \eeq
 with  $s= \frac{1-\gamma}{2}$, provided $\gamma<1$, and \beq\int_\ell \tilde A=0\eeq if $\gamma >1$.  Hence, the line integral  $A$ or $\tilde A$ cannot both yield integer values, the basic requirement for quantization. The same for $\tilde A$ applies to $a$. The only difference is  the role of $\gamma$ except in this case there is an interchange in the sense that 
\beq  {\mathrm Norm}\left( \int_\ell a \right)=\frac {\int_\ell A}{\Gamma(s+1)} \eeq
with $s= \frac{1-\gamma}{2}$, when $\gamma >1$ and
\beq \int _\ell a =0\eeq
when $\gamma <1$. The issue of (the lack of) quantization of the charge of the auxiliary field $a$ carries along a question of how the phase changes, as far as matter is concerned.   What is required to solve the problem of the phase is the appropriate covariant derivative $D_i \equiv \partial_i - i\frac{e}{\hbar}a_i$ in which the gauge field satisifying the ``local'' gauge group acts. Here only the spatial part of $\Box$ in the definition of the form $a$ is taken, and hence $a_i = (-\Delta)^{\frac{1-\gamma}{2}}A_i$. 
\noindent
  Choosing $A_0=0$, we reduce the Schr\"{o}dinger equation accordingly,
\begin{equation} \label{eq:schro_frac_A}
\bigg(-\frac{\hbar^2}{2m}(\partial_i - i\frac{e}{\hbar}a_i)^2+V\bigg)\psi = i\hbar\partial_t\psi.
\end{equation}
To derive the Aharonov-Bohm (AB) phase, let us consider a particle confined to the $x,y$ plane with a fractional magnetic field applied along the $z$ axis. Assume a particle can move from point $\vec r{_i}$ to $\vec r{_f}$ along path $\gamma_1$ (with wave function $\psi_1$) and along  path $\gamma_2$ (with wave function $\psi_2$). The total wave function at the point $\vec r{_f}$ at zero fractional magnetic field ($a_i = 0$) is $\psi = \psi_1 + \psi_2$. When the fractional magnetic field is turned on, the total wave function at $\vec r{_f}$ changes to
\begin{eqnarray}
\psi(\vec r_f,t) &=& e^{i\frac{e}{\hbar}\int_{\gamma_1}\vec a(\vec r)\cdot d \vec l}\psi_1(\vec r{_f},t) \nonumber \\
&& + e^{i\frac{e}{\hbar}\int_{\gamma_2}\vec a(\vec r)\cdot d \vec l}\psi_2(\vec r{_f},t) \nonumber\\
&=&  C \bigg( \psi_1(\vec r{_f},t) + e^{i\frac{e}{\hbar}\oint \vec a(\vec r)\cdot d \vec l} \psi_2(\vec r{_f},t) \bigg). \nonumber \\
\end{eqnarray}
Here $C$ is an over all phase factor $= e^{i\frac{e}{\hbar}\int_{\gamma_1} \vec a(\vec r)\cdot d \vec{l}}$. The phase difference between the two paths due to the gauge field is
\begin{equation} \label{eq:frac_AB_phase}
\Delta \phi = \frac{e}{\hbar}\oint \vec a(\vec r)\cdot d \vec l.
\end{equation}
In the strange metal, we posit that the current carrying degrees of freedom which emerge in the infrared couple to the fractional electromagnetic fields.  By definition, the propagating degrees of freedom are weakly interacting thereby warranting the Schr\"{o}dinger propagator approach we have been adopted here.

The AB phase\cite{ABnew} shift for the rotationally invariant definition is easily derived using the momentum-space formulation of the fractional Laplacian.  The result,
\begin{widetext}  \begin{eqnarray} \label{eq:phase_circle}
\Delta \phi_{\rm D} = \frac{e}{\hbar}\pi r^2 {_\alpha}B R^{2\alpha-2} \left(\frac{2^{2-2\alpha}\Gamma(2-\alpha)}{\Gamma(\alpha)} {_2}F_1(1-\alpha,2-\alpha,2;\frac{r^2}{R^2})\right).
\end{eqnarray} \end{widetext}
involves the standard result, $\pi r^2 B$ multiplied by a quantity that depends on the total outer radius of the sample such that the total quantity is dimensionless.  Here  $_2F_1(a,b;c;z)$ is a hypergeometric function and the terms in the parenthesis reduce to unity in the limit $\alpha \rightarrow 1$.  This is the key experimental prediction of the fractional formulation of electricity and magnetism:  the  flux depends on the outer radius.  This stems from the non-local nature of the underlying theory.  
  
\begin{figure}
	\includegraphics[scale=0.5]{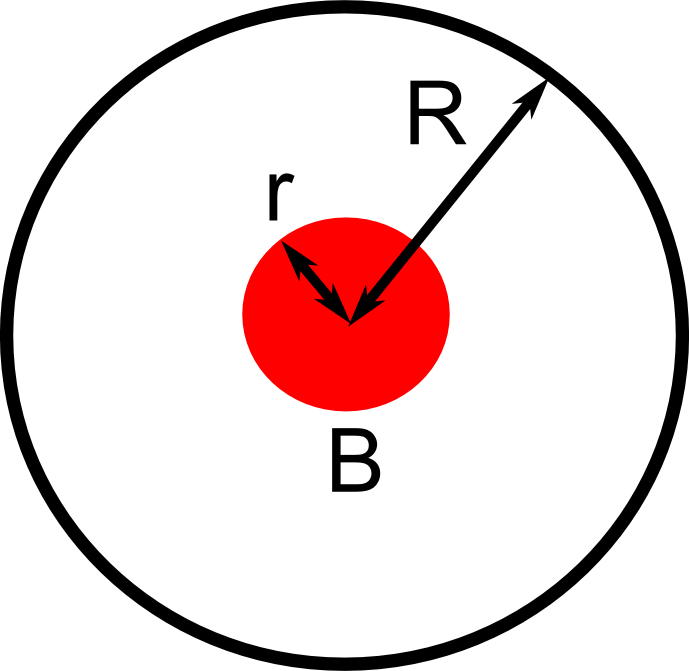} 
    \caption{Disk geometry for AB phase calculation.  The fractional magnetic field pierces the disk in a small region of radius, $r$. Reprinted from Euro. Phys. Lett. 121 (2018) 27003. }  \label{fig:disk}
\end{figure}

\section{Conclusion}

We have formulated here a theory of electricity and magnetism in which the underlying gauge field and current have non-traditional dimensions.  Such a formulation necessitates replacing the standard gauge invariant condition, $A\rightarrow A+d\Lambda$, with one that preserves simultaneously the 1-form nature of $A$ and the coordinate-independence of the underlying theory.  Given that gauge symmetries arise from local differential operators, this seriously restricts the viable options for an alternative formulation.  The insight from N\"other's second theorem and the zero-eigenvalue of Eq. (\ref{mgen}) imply that the unique equation that must be satisfied is $[d,\hat Y]=0$.  The only non-trivial solution for $\hat Y$ is the fractional Laplacian. As proven here, this provides an exact solution to the holographic dilatonic models.  As such models yield the same boundary or horizon action as those in which the bulk contains a massive gauge coupling along the radial direction only, the underlying mechanism for anomalous dimensions appears to be breaking of $U(1)$ symmetry down to $\mathbb Z_2$.   As a consequence, our treatment  unifies two key examples where anomalous dimensions to gauge fields and their associated currents occur, the other case being the Pippard\cite{pippardref} treatment of the Meissner effect.  
To reiterate, the anomalous dimension for the charge in QED 
for d = 2,3 and also in more general settings\cite{wise} are not  examples
of the physics we have treated here as $[qA]=1$.  At play in the examples we have constructed here is the extra redundancy delineated by N\"other in her Second Theorem.  It is on this principle that fractional electricity and magnetism hinges.  

Two key predictions of fractional electricity and magnetism is the lack of quantization of charge and an anomalous power for the skin depth in AC conductivity measurements. The strange metal offers a platform for falsifying both of these predictions as both experiments can be carried out in the normal state of the cuprates.

PWP thanks  E. Witten for encouragement and his characteristically level-headed remarks and  M. Stone and R. Gianetta for helpful suggestions regarding the Pippard kernel.  We acknowledge support from the Center for Emergent
Superconductivity, a DOE Energy Frontier Research Center, Grant No. DE-AC0298CH1088.  We also thank the NSF DMR-1461952 for partial funding of this project.  

\section{Appendix A: Breakdown of Charge Quantization}

In this appendix, we prove the lack of quantization of the charge for fractional gauge fields.  The identity, Eq. (\ref{alternative}),  only holds in a suitable renormalized sense that we elucidate in what follows, namely we define
\begin{footnote}{Lest this should be too obscure, let us remark that $\Delta ^{-s} A= \lim _{\Lambda \to +\infty} \frac{1}{\Gamma (s)}\int_0^{\Lambda} \, e^{-t\Delta}A \frac{dt}{t^{1-s}}$ and that $\Delta ^{-s}1=  \lim _{\Lambda \to +\infty}\frac{1}{\Gamma (s)}\int_0^{\Lambda} \, \frac{dt}{t^{1-s}}$}\end{footnote},
 \begin{eqnarray}  &{\mathrm Norm}\left( \int_\ell \tilde A\right)\\= &\lim_{\Lambda \to +\infty} \frac{1}{\Lambda ^s} \left( \frac{1} {\Gamma (s)} \int_0^{\Lambda} \, \int_\ell  e^{t\Delta} A \frac{dt}{t^{1-s}} \right).\nonumber \end{eqnarray}
This is basically a renormalization of IR divergences and mathematically related to the fact that $(-\Delta)^{-s} 1$ is infinite on the whole $\mathbb R^n$ for $s<0$. In fact, it is correct to think of the renormalization $\rm Norm$ in the fashion,
 \beq \label{norm-fraclap}{\mathrm Norm}\left( \int_\ell \tilde A\right)= \frac{1}{\Gamma (s+1)}\; \frac{ \int_\ell \tilde A}{\Delta ^{-s}1}.
 \eeq
 Alternatively, in order to avoid renormalization issues, we assume that the EM-fields are defined in a compact domain 
 $\Omega$ which we assume to be a big enough ball centered at the origin: $B_R(0)$.  Given this, we define
 \beq \label{alternative-domain}
 \int_\ell \tilde A=\zeta_{ \int_\ell A}\left(\frac{(\gamma-1)}{2}\right)
 \eeq
 provided $\gamma<1$, and 
 \beq\int_\ell \tilde A=0\eeq
  if $\gamma >1$ with $s = \frac{1-\gamma}{2}$. Here $\zeta _\alpha (s)$ is a zeta-like function to be defined below.
 
 To proceed, we switch to Euclidean signature and use $\Delta$ for the Hodge Laplacian $dd^*+d^*d$. 
We also recall the following standard facts:
\begin{itemize}
\item (Functoriality) for any smooth map of manifold $\iota :\Sigma \to M$ and for any $p$-form $\alpha$ on $M$, we have $\iota ^* d_M \alpha= d_\Sigma \iota ^*\alpha$ and the analogous formula with $d$ replaced by $d^*$ (here we make use of a subindex in the differential, to emphasize which manifold it is taken on);
\item (Integration by parts) On a {\it closed} (i.e., compact and without boundary) manifold $\Sigma$, $\int _\Sigma \,d_\Sigma d_\Sigma ^* \alpha \wedge \beta= \int _\Sigma \, d_\Sigma ^* \alpha \wedge d_\Sigma ^*\beta$ and the analogous formula with $d_\Sigma ^*$ replacing $d_\Sigma $;
\item  (Corollary of Integration by parts) On a closed manifold $\Sigma$, $\int _\Sigma  \Delta_\Sigma \omega =0$ (since $\Delta_\Sigma = d_\Sigma d_\Sigma ^*  + d_\Sigma ^* d_\Sigma $).
\end{itemize}
Now observe that, if $\gamma <1$, by eq. (\ref{fractionalintegral})
 \beq
\Delta ^{\frac{(\gamma-1)}{2}} A = \frac{1}{\Gamma(s)}\int_0^{+\infty} e^{-t\Delta} A \frac{dt}{t^{1-s}},
\eeq
with $s=-{\frac{(\gamma-1)}{2}}$.

Our first observation is that $\int _\ell \, e^{-t\Delta} A$ must be constant (hence equal to its initial value $\int _\ell A$).
This follows because the heat {\it semigroup} $e^{-t \Delta} A$ on forms is defined by requiring that $\beta=e^{-t \Delta} A$ be the solution to the diffusion equation
\beq
\frac{\partial} {\partial t}  \beta + \Delta \beta = 0
\eeq
with initial condition $\beta (x,0) = A$ and therefore $\frac{d}{dt} \int_\ell \beta= \int _\ell \frac{\partial \beta}{\partial t} = -\int_\ell \Delta \beta=0$, where we have made use of functonality in the standard facts stated above to conclude that $\iota ^* \Delta = \Delta _\ell$ (where $\iota:\ell \to \mathbb R^n$ is the inclusion of the loop) and that, since $\ell$ is a closed submanifold, that $\int _\ell \Delta _\ell \beta=0$. Therefore, $\int_\ell \beta$ is constant and equal its initial value $\int _\ell A$. 
This yields, setting $s= \frac{(\gamma-1)}{2}$, 
\beq {\mathrm Norm}\left( \int _\ell \tilde A\right)&=& \lim _{\Lambda \to +\infty} \int_\ell\left(\frac{1}{\Gamma(s) \Lambda ^s}\int_0^{\Lambda} e^{-t\Delta} A \frac{dt}{t^{1-s}}\right)\nonumber \\
&=& \lim _{\Lambda \to +\infty} \frac{1}{\Gamma(s)\Lambda ^s}\int_0^{\Lambda} \; \left(\int _\ell A\right)  \frac{dt}{t^{1-s}}\nonumber \\
&=&  \int _\ell A \; \lim _{\Lambda \to +\infty} \frac{1}{\Gamma(s)\Lambda ^s}\int_0^{\Lambda} \;  \frac{dt}{t^{1-s}}\nonumber \\
&=&\frac{1}{\Gamma(s+1)} \int _\ell A
\eeq
which is what claimed in Eq. (\ref{alternative}), having used the fact that for any number $\Lambda$ the identity $ \frac{1}{\Gamma(s)}\int_0^{+\Lambda} \;   \frac{dt}{t^{1-s}}= \frac{\Lambda ^s}{s\Gamma(s)}= \frac{\Lambda ^s}{\Gamma(s+1)}$ holds.
 Similarly, if $\gamma >1$, $\int _\ell \tilde A=0$.
 
%
 
 As for the case of the finite domain $\Omega$, we just observe that in that case, if $A_i$ is a {\it complete} basis of (1-form) eigenvalues of the Laplacian $\Delta A_i+\lambda _i A_i =0$ on $\Omega$ (here $A_i$ is not the i-th component of a form $A$, but rather an index related to the spectral decomposition), then the heat kernel is $e^{-t\Delta} A= \int_\Omega \, H(x,y,t) A$ with $ H(x,y,t) = \sum _{i=1}^{\infty} e^{-\lambda _i t} A_i (x)  A_i(y)$ so that
 \beq e^{-t\Delta} A=  \sum _{i=1}^{\infty} e^{-\lambda _i t}\alpha _i A_i(x)\eeq
 where $\alpha _i =\int _{\Omega} A\wedge \star A_i$ and therefore 
\begin{eqnarray} \Delta ^{-s} A &=  \frac{1}{\Gamma(s)}\int_0^{+\infty} e^{-t\Delta} A \frac{dt}{t^{1-s}}\nonumber \\&= \sum _{i=1}^{\infty} \lambda _i ^{-s}\alpha _i A_i(x)\end{eqnarray}
 having used that $ \frac{1}{\Gamma(s)}\, \int_0^{\infty} e^{-\lambda t} \frac{dt}{t^{1-s}}= \lambda ^{-s}$. Therefore,
 \beq \int_\ell \Delta ^{-s} A=  \sum _{i=1}^{\infty} \lambda _i ^{-s}\alpha _i \int_\ell A_i(x)\eeq
 which is what we wanted to prove, setting
 \beq \zeta_{ \int_\ell A}\left(\frac{(\gamma-1)}{2}\right)=  \sum _{i=1}^{\infty} \lambda _i ^{-s}\alpha _i \int_\ell A_i(x).\eeq

\section{Appendix B: Witten's Argument}
Here we want to report a simple and beautiful argument  due to E. Witten, referred to us in private conversation with the third author, which reduces the CS mechanism for forms to the one for functions, when the background spacetime is $2+1$-dimensional.

We therefore restrict to the case in which the background space (of Euclidean signature) is ~3-dimensional and exploit heavily the fact that the Hodge star operator in this case permutes ~2-forms into ~1-forms. Naturally this works only in dimension ~3. 
We will make henceforth the assumption that we are on a ~3-dimenaional manifold $M$ with boundary (in the CS mechanism, this is just $\mathbb R ^2\times \mathbb R_+$) such that $H^1(M , \mathbb R)=0$ or that we are on a patch $U\subset \Sigma$ containing a portion of the boudnary of $\Sigma$, for which $H^1(U,  \mathbb R)=0$. This condition is equivalent to requiring that closed forms are exact, by definition.
To put into formal mathematics what we envisaged earlier, we have that the star operator is an isomorphism $\star :\Omega^3\to \Omega ^1$.

We write equation (\ref{formsCS}) for $A$ (omitting the boundary data for simplicity of notation) as
\beq\label{1-formsCSdim2}
d^\star (y^adA)=0 \qquad \text{ and } \qquad  d(y^ad^\star A)=0
 \eeq
 Setting $C= y^a \star dA$, which is a ~1-form, we obtain that $dC=0$ and by Poncar\'e's Lemma, $C= d\phi$\begin{footnote}{Due to the cohomological assumptions on the manifold (or its patch) we can write that $y^a\star dA= d\phi$ for some smooth function $\phi$, because under these assumptions Poncar\'e's Lemma says that a closed ~1-form is exact .}\end{footnote}. Therefore $dA= y^{-a} \star d\phi$ and since $dA$ is closed (more generally, due to the Bianchi identity for the curvature $F_A$), we conclude $ d( y^{-a} \star d\phi)=0$
 and the therefore (using that $d^\star= \star d\star$)
 \beq 
 d^\star( y^{-a}  d\phi)=0.
 \eeq
This is a Caffarelli-Silvestre equation for $\phi$. 

\end{document}